\documentclass[11pt, a4paper]{openjournal}

\pdfoutput=1

\usepackage{amsmath}
\usepackage{graphicx}

\usepackage[pdftex, colorlinks=true, linkcolor=black, urlcolor=black, citecolor=blue]{hyperref}

\usepackage[english]{babel}
\usepackage[utf8]{inputenc}
\usepackage{enumerate}
\usepackage{amsbsy}
\usepackage{amssymb} 

\usepackage{amsfonts}
\usepackage{pstricks}
\usepackage{color}
\usepackage{setspace}

\newcommand{\bea}{\begin{eqnarray}} \newcommand{\eea}{\end{eqnarray}}
\newcommand{\el}{\nonumber \\}
\newcommand{\re}[1]{(\ref{#1})}

\newcommand{\pat}{\partial}

\renewcommand{\sec}[1]{section \ref{#1}}

\newcommand{\para}{\paragraph}

\renewcommand{\a}{\alpha}
\renewcommand{\b}{\beta}
\renewcommand{\c}{\gamma}
\renewcommand{\d}{\delta}

\newcommand{\ha}{\frac{1}{2}}

\newcommand{\rmd}{\mathrm{d}}

\newcommand{\ie}{i.e.\ }

\newcommand{\Mpl}{M_{{}_{\mathrm{Pl}}}}

\begin{document}

\begin{flushleft}
	\hfill		 HIP-2018-27/TH \\
\end{flushleft}

\title{Higgs inflation in the Palatini formulation with kinetic terms for the metric}

\author{Syksy R\"{a}s\"{a}nen}
\email{syksy.rasanen@iki.fi}
\address{University of Helsinki, Department of Physics and Helsinki Institute of Physics,\\ P.O. Box 64, FIN-00014 University of Helsinki, Finland \\ Birzeit University, Department of Physics \\ P.O. Box 14, Birzeit, West Bank, Palestine}

\begin{abstract}
\noindent
We consider scalar field inflation in the Palatini formulation of general relativity. The covariant derivative of the metric is then non-zero. 
From the effective theory point of view it should couple to other fields.
We write down the most general couplings between it and a scalar field that are quadratic in derivatives. We consider both the case when the torsion is determined by the field equations and the case when it is assumed to be zero a priori.
We find the metric derivative terms can significantly modify inflationary predictions.
We specialise to Higgs inflation and terms of only up to dimension 4. Transforming to the Einstein frame, we show that by tuning the coefficients of the new terms, we can generate various effective inflationary potentials, including quadratic, hilltop-type, $\alpha$-attractor and inflection point. Some of these can give inflation in agreement with observations, including with a large tensor-to-scalar ratio, even if the non-minimal coupling is zero.
\end{abstract}

\maketitle

\setcounter{tocdepth}{2}

\setcounter{secnumdepth}{3}

\section{Introduction} \label{sec:intro}

Inflation is the most successful scenario for the primordial universe. It alleviates the homogeneity and isotropy problem \cite{Vachaspati:1998dy, Trodden:1999wc, Ellis:1999sx, ellis2000, East:2015ggf}, explains spatial flatness and has (in its simplest variants) predicted that primordial perturbations are mostly adiabatic, close to scale-invariant, highly Gaussian, predominantly scalar and statistically homogeneous and isotropic to a high degree \cite{Starobinsky:1980te, Kazanas:1980tx, Guth:1981, Sato:1981, Mukhanov:1981xt, Linde:1981mu, Albrecht:1982wi, Hawking:1981fz, Chibisov:1982nx, Hawking:1982cz, Guth:1982ec, Starobinsky:1982ee, Sasaki:1986hm, Mukhanov:1988jd}. This is in excellent agreement with observations \cite{Planck2018}. Higgs inflation \cite{Bezrukov:2007ep} (for reviews, see \cite{Bezrukov:2013fka, Bezrukov:2015, Rubio:2018}) is an attractive model where inflation is driven by the only fundamental scalar field in the Standard Model (SM) of particle physics. This is not possible with the SM action alone, as the Higgs potential is not sufficiently flat \cite{Isidori:2007vm, Hamada:2013mya, Fairbairn:2014nxa}.\footnote{False vacuum inflation is a possibility, but requires physics beyond the SM for graceful exit \cite{Fairbairn:2014nxa, Masina:2011aa, Masina:2011un, Masina:2012yd, Masina:2014yga, Notari:2014noa, Iacobellis:2016}.} However, including a non-minimal coupling between the Higgs field and gravity decelerates the field sufficiently. (In the Einstein frame the potential becomes exponentially flat, in the Jordan frame there is extra friction.) The predictions of the classical action agree well with observations \cite{Planck2018}. Loop corrections complicate the picture \cite{Espinosa:2007qp, Barvinsky:2008ia, DeSimone:2008ei, Bezrukov:2008ej, Burgess:2009ea, Bezrukov:2009db, Barvinsky:2009ii, Popa:2010xc, Bezrukov:2010jz, Bezrukov:2012sa, Allison:2013uaa, Salvio:2013rja, Shaposhnikov:2013ira, George:2013iia, Calmet:2013hia, Moss:2014, Hamada:2014iga, Bezrukov:2014bra, Bezrukov:2014ipa, Rubio:2015zia, George:2015nza, Saltas:2015vsc, Fumagalli:2016lls, Bezrukov:2017dyv, Enckell:2016xse, Rasanen:2017, Enckell:2018a, Jinno:2017a, Jinno:2017b, Markkanen:2017, Masina:2018}, and can change the qualitative behaviour to produce near-inflection point inflation \cite{Allison:2013uaa, Bezrukov:2014bra, Hamada:2014iga, Bezrukov:2014ipa, Rubio:2015zia, Fumagalli:2016lls, Enckell:2016xse, Bezrukov:2017dyv, Rasanen:2017, Masina:2018ejw, Salvio:2017oyf, Ezquiaga:2017fvi, Rasanen:2018a}, hilltop inflation \cite{Fumagalli:2016lls, Rasanen:2017, Enckell:2018a}, hillclimbing inflation \cite{Jinno:2017a, Jinno:2017b} and generate higher order curvature terms in the action that  change the evolution \cite{Barbon:2015, Salvio:2015kka, Salvio:2017oyf, Kaneda:2015jma, Calmet:2016fsr, Wang:2017fuy, Ema:2017rqn, Pi:2017gih, He:2018gyf, Gorbunov:2018llf, Ghilencea:2018rqg, Wang:2018, Enckell:2018b, Antoniadis:2018ywb, Gundhi:2018wyz, Antoniadis:2018yfq}. The issue of unitarity also remains unsettled \cite{Barbon:2009ya, Burgess:2009ea, Hertzberg:2010dc, Bauer:2010, Bezrukov:2010jz, Bezrukov:2011sz, Calmet:2013hia, Weenink:2010rr, Prokopec:2012ug, Xianyu:2013, Prokopec:2014iya, Ren:2014, Escriva:2016cwl, Gorbunov:2018llf}.

Another complication is the dependence of the predictions on the choice of the gravitational degrees of freedom. (On different formulations of general relativity and related theories, see \cite{Sandberg:1975, Hehl:1976, Hehl:1978, Papapetrou:1978, Hehl:1981, Percacci:1991, Rovelli:1991, Nester:1998, Percacci:2009, Krasnov:2017, Adak:2005, Gielen:2018, Jarv:2018, BeltranJimenez:2018, Koivisto:2018, BeltranJimenez:2017, Conroy:2017, Harko:2018}.) In the metric formulation of general relativity, the only independent gravitational degree of freedom is the metric, and all geometric properties of the manifold are derived from it. In particular, the connection is the Levi--Civita connection that is uniquely determined by the metric. However, the metric and the connection describe different aspects of the manifold. The metric defines distances in spacetime and dot products between vectors in the tangent space, whereas the connection sets straight lines and gives the derivatives of tensor fields. In the Palatini formulation, the metric and the connection are taken to be independent degrees of freedom.\footnote{The Palatini formulation is somewhat of a misnomer, as it was developed by Einstein \cite{einstein1925, ferraris1982}. It is also called the first order formulation, as only first derivatives appear in the gravitational action, unlike in the metric formulation (also referred to as the Hilbert formulation). Metric-affine is perhaps a more descriptive term, but it is often used to refer to the case when the general connection is used in both the gravitational and the matter action, with the name Palatini restricted to the case when the general connection is reserved for the gravitational sector, and the Levi--Civita connection is used in the matter sector \cite{Sotiriou:2006, Sotiriou:2008, Olmo:2011}.} In the metric formulation, the only tensors that can be constructed from the gravitational degrees of freedom are the metric, the Levi--Civita tensor and the Riemann tensor. In the Palatini formulation, applying the covariant derivative to the metric gives a new tensor, the non-metricity tensor $Q_{\c\a\b}=\nabla_\c g_{\a\b}$. If the connection is not symmetric, its antisymmetric part, the torsion, is a third gravitational tensor. A number of new gravitational terms can thus be added to the action.

In the Palatini case, the equations of motion do not determine the metric and the connection uniquely for all actions \cite{Sandberg:1975, Buchdahl:1960, Hehl:1978, Papapetrou:1978, Hehl:1981, Cotsakis:1997, Querella:1999}. This happens for the Einstein--Hilbert action if the connection is not taken to be symmetric. Rather than a drawback of the Palatini formulation, this can be seen as a constraint on the action, showing that extra terms are needed to break the projective invariance that is the reason for the system being underdetermined. An alternative viewpoint is to use Lagrange multipliers to impose the desired properties of the connection \cite{Tsamparlis:1977, Hehl:1978, Hehl:1981, Cotsakis:1997, Querella:1999, Iglesias:2007, Jarv:2018, BeltranJimenez:2017, BeltranJimenez:2018}, or simply assume them a priori. However, there are actions, for example the $R^2$ action, for which the equations of motion do not determine the degrees of freedom even if the connection is symmetric.

For the Einstein--Hilbert action plus a matter action that is coupled only to the metric, not to the connection, the equation of motion for a symmetric connection gives the Levi--Civita connection, so the metric and the Palatini formulation are physically equivalent. When the gravitational action is more complicated \cite{Harko:2018, Buchdahl:1960, Buchdahl:1970, ShahidSaless:1987, Flanagan:2003a, Flanagan:2003b, Sotiriou:2006, Sotiriou:2008, Olmo:2011, Cuzinatto:2016, Cuzinatto:2018, Borunda:2008, Querella:1999, Cotsakis:1997, Jarv:2018, Conroy:2017, Li:2007, Li:2008, Exirifard:2007, BeltranJimenez:2018, BeltranJimenez:2017, Koivisto:2018, Enckell:2018b, Antoniadis:2018ywb, Kijowski:2016, Afonso:2017, Afonso:2018a, Afonso:2018c, Iosifidis:2018zjj} or the matter action couples to the connection \cite{Lindstrom:1976a, Lindstrom:1976b, Bergh:1981, Bauer:2008, Bauer:2010, Koivisto:2005, Rasanen:2017, Enckell:2018a, Markkanen:2017, Kozak:2018, Jarv:2017, Iosifidis:2018zwo}, this is no longer true. In particular, this is the case when a scalar field is coupled directly to the Ricci scalar \cite{Lindstrom:1976a, Lindstrom:1976b, Bergh:1981, Bauer:2008, Bauer:2010, Koivisto:2005, Rasanen:2017, Enckell:2018a, Markkanen:2017, Jarv:2017, Kozak:2018, Tenkanen:2017, Carrilho:2018}. In Higgs inflation, this means that the same term that enables inflation breaks the equivalence between the metric and the Palatini formulation \cite{Bauer:2008, Bauer:2010, Rasanen:2017, Enckell:2018a, Markkanen:2017, Rasanen:2018a}.

If we combine general relativity and the SM of particle physics, in the metric case the only new dimension 4 term that could not have been written down with either theory alone is the non-minimal coupling $\xi |H|^2 R$, where $H$ is the Higgs doublet. In the Palatini case, the non-metricity tensor $Q_{\a\b\c}$ can couple to SM fields, in particular to the gradient of the Higgs field. Both the non-minimal coupling and the kinetic mixing between the metric and the Higgs then act as sources for $Q_{\a\b\c}$.

Assuming that the connection enters only via the Riemann tensor and the covariant derivative, we write down the most general classical action built from the metric, the connection and a scalar field that involves only up to two derivatives and no boundary terms, excepting the Holst term and Riemann tensor squared terms. We will consider separately the cases when the connection is assumed to be symmetric a priori, and when the antisymmetric part is left free to be determined by the field equations. In \sec{sec:kin} we present the action, derive the equations of motion and transform to the Einstein frame. In \sec{sec:Higgs} we specialise to SM Higgs inflation and terms of only up to dimension 4 and study the effect of the new terms on the effective Higgs potential. In \sec{sec:disc} we discuss our results and open issues, and in \sec{sec:conc} we summarise our findings.

\section{Inflation with kinetic terms for the metric} \label{sec:kin}

\subsection{Action and degrees of freedom}

\para{The degrees of freedom.}

We will consider three independent degrees of freedom: scalar field $h$, symmetric metric $g_{\a\b}$ and connection $\Gamma^{\c}_{\a\b}$. We will comment on fermions and gauge fields in \sec{sec:disc}. We decompose the connection as
\bea \label{Gamma}
  \Gamma^{\c}_{\a\b} = \mathring\Gamma^{\c}_{\a\b} + L^{\c}_{\ \a\b} \ ,
\eea
where $\mathring\Gamma_{\a\b}^{\c}$ is the Levi--Civita connection defined by the metric $g_{\a\b}$. As the difference of two connections, $L^{\c}_{\ \a\b}$ is a tensor, known as the deformation tensor. We can decompose $L^{\c}_{\ \a\b}$ into parts that depend on torsion tensor and on the non-metricity tensor defined as, respectively,
 \bea \label{TQ}
  T^\c_{\ \a\b} &\equiv& 2 \Gamma^{\c}_{[\a\b]} \ , \qquad  Q_{\c\a\b} \equiv \nabla_\c g_{\a\b} \ .
\eea
Note that $Q_{\c\a\b}=Q_{\c(\a\b)}$ and $\nabla_\c g^{\a\b}=-Q_{\c}^{\ \a\b}$. We also define $Q^\c\equiv g_{\a\b} Q^{\c\a\b}$, $\hat Q^\b\equiv g_{\a\c} Q^{\a\b\c}$ and $T^\b\equiv g_{\a\c} T^{\a\b\c}$. Denoting the covariant derivative defined by the Levi--Civita connection by $\mathring\nabla$, we have for an arbitrary vector $A^\a$
\bea \label{nabla}
  \nabla_\b A^\a = \mathring\nabla_\b A^\a + L^{\a}_{\ \b\c} A^\c \ .
\eea
As $\mathring\nabla_\c g_{\a\b}=0$ by definition, we get
\bea \label{Q}
  Q_{\c\a\b} &=& \nabla_\c g_{\a\b} = - L_{\a\c\b} - L_{\b\c\a} \ .
\eea
We can write $L_{\a\b\c}$ as
\bea \label{L}
  L_{\a\b\c} = \frac{1}{2} \left( Q_{\a\b\c}  - Q_{\c\a\b} - Q_{\b\a\c} \right) + K_{\a\b\c}  \ ,
\eea
where $K^{\c\ \ \b}_{\ \ \a}=K^{[\c\ \ \b]}_{\ \ \a}$ is the contortion tensor, given in terms of the torsion as
\bea \label{K}
  K_{\a\b\c} = \ha ( T_{\a\b\c} + T_{\b\a\c} + T_{\c\a\b} ) \ .
\eea
We will discuss separately the case when no symmetry assumptions are made about the connection and the case when the torsion is taken to be zero a priori (\ie $L^{\c}_{\ \a\b}$ is symmetric).

\para{The action.}

We want write a bulk action from which equations of motion can derived without having to add boundary terms to cancel terms arising in the variation. (In the metric formulation the York--Gibbons--Hawking boundary term is needed to cancel a boundary term involving the variation of the first derivative of the metric \cite{York:1972, Gibbons:1977}.) We will include $L_{\a\b\c}$ only as part of the connection, so it appears only in covariant derivatives and in the Riemann tensor. We cannot get kinetic terms for $L_{\a\b\c}$ from covariant derivatives of the metric, because they would involve the second derivative of $g_{\a\b}$, leading to terms proportional to the variation of the first derivative on the boundary, which does not vanish. Without kinetic terms, the connection reduces to an auxiliary variable, so the Palatini formulation does not introduce new degrees of freedom compared to the metric case, it only changes the relation between existing degrees of freedom.\footnote{In the case when the equation of motion of the connection has no derivatives, the Palatini formulation can be viewed as the low-energy limit of a theory with a connection field (more precisely, $L_{\a\b\c}$) that has high mass \cite{Percacci:1991, Percacci:2009, Krasnov:2017}. For related considerations of confinement of the spin connection, see \cite{Donoghue:2016a, Donoghue:2016b}.}
We will not consider terms non-linear in the Riemann tensor, nor the parity-violating Holst term. We will be mainly interested in terms quadratic in the derivatives and only up to dimension 4, in which case such Riemann tensor terms do not couple to the scalar field. Also, unless the coefficients of the second order Riemann terms are large, they are not important for inflation in the observed range of e-folds, because the curvature is much smaller than the Planck scale, $R/\Mpl^2=2.7\times10^{-9} r/0.07$ \cite{Planck2018}, where $r<0.07$ is the tensor-to-scalar ratio. The effect of Ricci scalar squared terms in non-minimally coupled scalar field inflation in the Palatini formulation has been studied in \cite{Enckell:2018b, Antoniadis:2018ywb}. 

The most general action built from scalar field $h$, metric $g_{\a\b}$ and connection $\Gamma^{\c}_{\a\b}$ containing only up to quadratic derivative terms is then
\bea
  \label{actionJg} S &=& \int\rmd^4 x \sqrt{-g} \left[ \ha F(h) g^{\a\b} R_{\a\b}(\Gamma, \pat\Gamma) - \ha K(h) g^{\a\b} \nabla_\a h \nabla_\b h - V(h) \right. \el
   && + A_1(h) \nabla_\a h \nabla_\b g^{\b\a} + A_2(h) g^{\a\b} g_{\c\d} \nabla_\a h \nabla_\b g^{\c\d} \el
   && + B_1(h) g^{\a\b} g_{\c\d} g_{\epsilon\eta} \nabla_\a g^{\c\epsilon} \nabla_\b g^{\d\eta} + B_2(h) g_{\c\d} \nabla_\a g^{\b\c} \nabla_\b g^{\a\d} + B_3(h) g_{\a\b} \nabla_\c g^{\c\a} \nabla_\d g^{\d\b} \el
  && + B_4(h) g^{\a\b} g_{\c\d} g_{\epsilon\eta} \nabla_\a g^{\c\d} \nabla_\b g^{\epsilon\eta} + B_5(h) g_{\c\d} \nabla_\a g^{\a\b} \nabla_\b g^{\c\d} \el
 && \left. + C(h) \epsilon^{\a\b}_{\ \ \ \c\d} g_{\epsilon\eta} \nabla_\a g^{\c\epsilon} \nabla_\b g^{\d\eta} \right] \\
  \label{actionJQ} &=& \int\rmd^4 x \sqrt{-g} \left[ \frac{1}{2} F(h) g^{\a\b} R_{\a\b}(\Gamma,\pat\Gamma) -\ha K(h) g^{\a\b} \nabla_\a h \nabla_\b h - V(h) \right. \el
   && - A_1(h) \nabla_\a h \hat Q^\a - A_2(h) \nabla_\a h Q^\a \el
   && + B_1(h) Q_{\c\a\b} Q^{\c\a\b} + B_2(h) Q_{\c\a\b} Q^{\b\c\a} + B_3(h) \hat Q_\a \hat Q^\a + B_4(h) Q_\a Q^\a + B_5(h) Q_\a \hat Q^\a
 \el
 && \left. + C(h) \epsilon^{\a\b\c\d} g^{\epsilon\eta} Q_{\a\c\epsilon} Q_{\b\d\eta} \right] \ ,
\eea
where $g$ is the determinant of $g_{\a\b}$, $R_{\a\b}=R^{\c}_{\ \, \a\c\b}$ is the Ricci tensor\footnote{In the Palatini case $R_{\a\b}$ is not in general symmetric, but the antisymmetric part does not contribute to the action or the equations of motion. The quantity $R_{\a\b}$ is not the only independent first contraction of the Riemann tensor \cite{Borunda:2008}. However, there is only one total contraction, \ie the Ricci scalar is unique.} and $\epsilon_{\a\b\c\d}$ is the Levi--Civita tensor.
If we did not exclude boundary terms, there would be more possible terms, including those involving the Riemann tensor built from the Levi--Civita connection alone and terms that mix it and the full Riemann tensor \cite{Flanagan:2003b, Li:2008}.
In the action, the covariant derivatives acting on $h$ are of course partial derivatives, and the connection that they will be paired with in the equations of motion is determined by the vanishing of boundary terms when varying the action. The kinetic mixing terms on second line of \re{actionJg} that couple derivatives of the scalar field and of the metric have been discussed in \cite{Bergh:1981, Burton:1997a, Kozak:2018, Iosifidis:2018zwo}, but do not seem to have been studied in the context of inflation before.\footnote{The term kinetic mixing may be slightly inappropriate because in the equations of motion the partial derivatives of the metric cancel with the Levi--Civita part of the connection, and only terms algebraic in $Q_{\a\b\c}$ remain.}
They will source $Q_{\a\b\c}$, which will in turn change the evolution of the scalar field.
The quadratic terms on the third and fourth lines have been considered in different contexts in \cite{Bergh:1981, Percacci:1991, Burton:1997b, Adak:2005, Percacci:2009, Conroy:2017, Jarv:2018, BeltranJimenez:2017, BeltranJimenez:2018, Koivisto:2018, Iosifidis:2018zwo}.
In \re{actionJQ} we have written the action in terms of $Q_{\a\b\c}$ to more clearly display the structure where we effectively have two extra vector fields of mass dimension one (in addition to the traceless part of $Q_{\a\b\c}$, which we have not separated out). However, note that we do not vary the action with respect to them, but with respect to $g_{\a\b}$ and $\Gamma^{\c}_{\a\b}$. The $Q_{\a\b\c}$ terms contribute to both variations. If we were to make the split \re{Gamma} already in the action before applying the variational principle, we would get boundary terms involving the variation of the first derivative of the metric, which would not vanish, just as in the metric case.
With only scalar field matter, the Levi--Civita term on the last line does not contribute to the equations of motion, so we drop it henceforth.

With the set of independent variables $\{h, g_{\a\b}, \Gamma^{\c}_{\a\b}\}$, the variation of the action reads
\bea \label{dS}
  \d S &=& \d h \frac{\d S}{\d h} + \d g_{\a\b} \frac{\d S}{\d g_{\a\b}} + \d \Gamma^{\c}_{\a\b} \frac{\d S}{\d \Gamma^{\c}_{\a\b}} \el
  &=& \d h \frac{\d S}{\d h} + \d g_{\a\b} \frac{\d S}{\d g_{\a\b}} + \d \Gamma^{\c}_{(\a\b)} \frac{\d S}{\d \Gamma^{\c}_{(\a\b)}} + p \d \Gamma^{\c}_{[\a\b]} \frac{\d S}{\d \Gamma^{\c}_{[\a\b]}} \ .
\eea
The first variation gives the scalar field equation of motion, the second gives the generalisation of the Einstein equation, and the third determines the connection in terms of the scalar field and the metric. On the second line we have split the variation with respect to the connection into the symmetric and the antisymmetric part. In the unconstrained case we have $p=1$; if the connection is taken to be symmetric a priori, we have $p=0$.

When varying the fields, it is important to note that the contraction of the covariant derivative with a vector does not reduce to a boundary term, because the connection is not Levi--Civita. Instead, \re{nabla} gives
\bea \label{boundary}
  \nabla_\a A^\a &=& \mathring\nabla_\a A^\a - \ha Q_\a A^\a - T_\a A^\a \ ,
\eea
where the Levi--Civita covariant derivative $\mathring\nabla_\a A^\a$ gives a boundary term as usual.

Often in the Palatini formulation the connection is taken to be symmetric a priori, \ie the torsion is set to zero by hand. However, if the connection is taken to be an independent degree of freedom, from the effective theory point of view is no reason to assume that its antisymmetric part is zero. For the Einstein--Hilbert action, the equations of motion only show that a linear combination of $Q_{\a\b\c}$ and $T_{\a\b\c}$ vanishes, a part is left undetermined, because of the the invariance of the action under the projective transformation $\Gamma^{\c}_{\a\b} \rightarrow \Gamma^{\c}_{\a\b} + \delta^\c_{\ \, \a} A_\b$, where $A_\b$ is an arbitrary vector \cite{Hehl:1978}. If we assume that either  $Q_{\a\b\c}$ or $T_{\a\b\c}$ vanishes, then it follows that the other one is zero as well.
This can be seen as a motivation to extend the Einstein--Hilbert action so that the equations of motion fully determine the evolution of the degrees of freedom, rather than a reason to set some of the degrees of freedom to zero by hand. We will see that the quadratic $Q_{\a\b\c}$ terms we have introduced break the projective invariance, and the equations motion fully determine the connection.

\subsection{Connection equation of motion} \label{sec:field}

\para{General case.}

Variation of the action with respect to $\Gamma^{\c}_{\a\b}$ gives, making the split \re{Gamma} in the equation of motion and taking into account $\mathring\nabla_\c g_{\a\b}=0$,
\bea \label{Gammaeq}
   && 8 B_1 Q_{\a\b\c} + 8 B_2 Q_{(\b\c)\a} + 4 g_{\a(\b} ( 2 B_3 \hat Q_{\c)} + B_5 Q_{\c)} ) + 4 g_{\b\c} ( 2 B_4 Q_\a + B_5 \hat Q_\a ) \el
   && - F Q_{\c\a\b} + F g_{\a\c} \hat Q_\b + F g_{\a[\b} Q_{\c]} + F T_{\a\b\c} + 2 F g_{\a[\b} T_{\c]} \el
   && = 4 A_1 g_{\a (\b} \pat_{\c)} h + 4 A_2 g_{\b\c} \pat_\a h - 2 F' g_{\a [\b} \pat_{\c]} h \ ,
\eea
where prime denotes derivative with respect to $h$. This shows how non-metricity and torsion are sourced by the derivative of the non-minimal coupling $F'$ and the kinetic mixings $A_i$.

The general solution of \re{Gammaeq} has the form
\bea \label{Qsol}
  Q_{\c\a\b} &=& q_1(h) g_{\a\b} \pat_\c h + 2 q_2(h) g_{\c(\a} \pat_{\b)} h \el
  T_{\a\b\c} &=& 2 t(h) g_{\a[\b} \pat_{\c]} h \ .
\eea
Inserting \re{Qsol} into \re{Gammaeq}, we get
\bea \label{qteq}
  ( a - 2 F t ) g_{\a\b} \pat_\c h + ( b + c + 2 F t ) g_{\c(\a} \pat_{\b)} h + p ( b - c + 2 F t ) g_{\c[\a} \pat_{\b]} h = 0 \ ,
\eea
where
\bea
  a &\equiv& - 2 A_1 + F' + ( 4 B_2 + 4 B_3 + 8 B_5 + F) q_1 + ( 8 B_1 + 4 B_2 + 20 B_3 + 4 B_5 + F) q_2 \el
  b &\equiv& - 2 A_1 - F' + ( 4 B_2 + 4 B_3 + 8 B_5 - F) q_1 + ( 8 B_1 + 4 B_2 + 20 B_3 + 4 B_5 + 3 F) q_2 \el
  c &\equiv& - 4 A_2 + ( 8 B_1 + 32 B_4 + 4 B_5 ) q_1 + ( 8 B_2 + 16 B_4 + 20 B_5 - F ) q_2 \ .
\eea

\para{Unconstrained case.}

In the unconstrained case ($p=1$), \re{qteq} gives $a=2 Ft, a+b=0, c=0$, leading to
\bea \label{qttor}
 q_1 &=& \frac{- A_1 ( 8 B_2 + 16 B_4 + 20 B_5 - F ) + 4 A_2 ( 4 B_1 + 2 B_2 + 10 B_3 + 2 B_5 + F )}{2 M} \el
 q_2 &=&\frac{ 2 A_1 ( 2 B_1 + 8 B_4 + B_5 ) - 4 A_2 ( B_2 + B_3 + 2 B_5 ) }{M} \el
 t &=& \{ - A_1 ( 8 B_1 + 8 B_2 + 48  B_4 + 24 B_5 - F ) + 4 A_2 ( 4 B_1 + 4 B_2 + 12 B_3 + 6 B_5 + F ) \el
 && + 2 [ 16 B_1^2 - 8 B_2^2 - 36 B_5^2 + 144 B_3 B_4 + 4 B_1 ( 2 B_2 + 10 B_3 + 16 B_4 + 4 B_5 ) \el
 && + B_2 ( - 8 B_3 +16 B_4 - 32 B_5 ) + ( 4 B_1 + B_2 + B_3  + 16 B_4  + 4 B_5 ) F ] F'/F \}/(4 M) \el
  \label{M} M &\equiv& 16 B_1^2 - 8 B_2^2 - 36 B_5^2 + 4 B_1 ( 2 B_2 + 10 B_3 + 16 B_4 + 4 B_5 ) + 144 B_3 B_4 \el
  && + B_2 ( - 8 B_3 + 16 B_4 - 32 B_5 ) + F ( 4 B_1 + B_2  + B_3 +16 B_4 + 4 B_5) \ .
\eea
If $M=0$, the equations do not in general have a solution. A particular case of this is $B_i=0$. If we put all $Q_{\a\b\c}$ terms to zero in the action, $A_i=B_i=0$, we get $2t-q_1=F'/F, q_2=0$, showing how only a particular linear combination of the torsion and non-metricity is determined. If we keep the quadratic terms but put the linear terms to zero, $A_i=0, B_i\neq0$, we get $q_1=q_2=0, t=F'/(2F)$. In other words, without the kinetic mixing terms, the non-minimal coupling generates torsion, but not non-metricity. If we also remove the non-minimal coupling (\ie put $F'=0$), the connection becomes Levi--Civita. So if were to only include the terms quadratic in $Q_{\a\b\c}$, they would successfully break the projective invariance and reduce the connection to Levi--Civita without otherwise affecting the equations of motion. However, when also $A_1, A_2$ and/or $F'$ is non-zero, the functional form of the quadratic terms is relevant, and they can have a significant effect on the evolution.

\para{Zero torsion.}

If torsion is taken to be zero a priori ($t=0, p=0$), \re{qteq} gives $a=0, b+c=0$, leading to
\bea \label{qtnotor}
 q_1 &=& [ - 2 A_1 ( 8 B_2 + 16 B_4 + 20 B_5 + F) + 4 A_2 ( 8 B_1 + 4 B_2 + 20 B_3 + 4 B_5 + F) \el
 && + (16 B_1 + 16 B_2 + 40 B_3 + 16 B_4 + 28 B_5 + 3 F) F' ]/N \el
 q_2 &=& 4 [  A_1 (4 B_1 + 16 B_4 + 2 B_5 - F) - A_2 ( 4 B_2 + 4 B_3 + 8 B_5 + F) \el
 && - ( 2 B_1 + 2 B_2 + 2 B_3 + 8 B_4 + 5 B_5 ) F' ]/N \el
 N &\equiv& 64 B_1^2 - 32 B_2^2 - 144 B_5^2 + 8 B_1 ( 4 B_2 + 20 B_3 + 32 B_4 + 8 B_5 ) - 4 B_2 ( 8 B_3 - 16 B_4 + 32 B_5 ) \el
 && + 576 B_3 B_4 - ( 8 B_1 + 20 B_2 + 44 B_3 - 16 B_4 + 32 B_5 + 3 F ) F \ .
\eea
If $N=0$, the equations do not in general have a solution. In contrast to the case with torsion, $N\neq0$ even if $B_i=0$. This corresponds to the fact that the projective invariance has been broken by setting the torsion to zero by hand. Here $F'$ sources non-metricity, unlike in the case with torsion, where it exclusively sources torsion and only $A_i$ act as sources for $Q_{\a\b\c}$. If we take $A_i=B_i=0$, we get $q_1=-F'/F, q_2=0$. This is the usual case of non-minimally coupled Higgs inflation in the Palatini formulation, studied originally in \cite{Bauer:2008}.

\para{Zero non-metricity.}

If we impose $Q_{\a\b\c}=0$ a priori instead of $T_{\a\b\c}=0$, the $Q_{\a\b\c}$ terms in the action obviously vanish, so we are left only with the non-minimal coupling $F$, and get the solution $t=F'/(2F)$. If $F'=0$, the torsion is zero, and we again recover the Levi--Civita connection, having set one of the degrees of freedom to zero by hand. It is straightforward to show that the torsion generated by $F'\neq0$ has exactly the same effect on the equations of motion of the scalar field and the metric as the non-zero $Q_{\a\b\c}$ in the above case when the torsion is set to zero and only the non-minimal coupling is relevant (\ie $A_i=B_i=0$). This has to be the case, because the non-minimal coupling can be eliminated via a conformal transformation of the metric, which knows nothing about the symmetries of the connection. So we see that the effect of the non-minimal coupling in usual Higgs inflation in the Palatini formulation can be mapped not only from the non-minimal coupling and non-metricity to the kinetic term or the potential, but also between the non-metricity and the torsion. (Mapping between non-metricity and torsion has also been studied in \cite{Afonso:2017, Iosifidis:2018zjj, Iosifidis:2018zwo}.) Let us now look at field transformations in more detail.

\subsection{Field transformations}

\para{General case.}

The equations of motion written in terms of the original variables $\{h, g_{\a\b}, \Gamma^{\c}_{\a\b}\}$ are rather complicated. Therefore, rather than solving them directly, as we did above for the connection, we simplify the problem by choosing suitable coordinates in field space. Without introducing higher order derivatives into the action, we can make the following field transformations:
\bea \label{trans}
  \label{transh} \!\!\!\!\!\!\!\!\!\!\! h &\rightarrow& \chi(h) \\
  \label{transg} \!\!\!\!\!\!\!\!\!\!\! g_{\a\b} &\rightarrow& \Omega(h)^{-1} g_{\a\b} \\
  \label{transG} \!\!\!\!\!\!\!\!\!\!\! \Gamma^{\c}_{\a\b} &\rightarrow& \Gamma^{\c}_{\a\b} + \Sigma^{\c}_{\ \a\b} = \Gamma^{\c}_{\a\b} + g^{\c\d} \left[ \Sigma_1(h) g_{\a\b} \pat_\d h + 2 \Sigma_2(h) g_{\d(\a} \pat_{\b)} h + 2 \Sigma_3(h) g_{\d[\a} \pat_{\b]} h \right] \, ,
\eea
so we have 5 free functions $\chi(h)$, $\Omega(h)$ and $\Sigma_i(h)$.\footnote{The term $\Sigma_4(h) \epsilon^{\c\d}_{\ \ \ \a\b} \pat_\d h$ could also be added to the torsion, but in our case where the torsion is generated by a scalar field its totally antisymmetric part is zero, so this term will not be needed.} The formulation of non-minimally coupled scalar field theories that is covariant with respect to transformations of the scalar field \re{transh} and the metric \re{transg} \cite{Jarv:2016sow, Karamitsos:2017elm, Karamitsos:2018lur, Karam:2017, Jarv:2014hma, Jarv:2015kga, Kuusk:2015dda} can be extended to also cover transformations of the connection \re{transG} \cite{Kozak:2018}. For more general mapping of gravitational degrees of freedom into matter degrees of freedom in the Palatini formulation, see \cite{Kijowski:2016, Afonso:2017, Afonso:2018a, Afonso:2018c}.

The scalar field redefinition \re{transh} simply gives
\bea
  \nabla_\a h &\rightarrow&  \frac{\rmd h}{\rmd\chi} \nabla_\a \chi \ .
\eea

Under the conformal transformation of the metric \re{transg}, we have
\bea \label{shiftg}
  g^{\a\b} &\rightarrow& \Omega g^{\a\b} \el
  \sqrt{-g} &\rightarrow& \Omega^{-2} \sqrt{-g} \el
  \nabla_\c g^{\a\b} &\rightarrow& \Omega ( \nabla_\c g^{\a\b} + g^{\a\b} \pat_\c \omega ) \ ,
\eea
where we have denoted $\omega\equiv\ln\Omega$. 

Under the translation of the connection \re{transG}, we have
\bea \label{shiftG}
  \nabla_\c g^{\a\b} &\rightarrow& \nabla_\c g^{\a\b} + 2 \Sigma^{(\a\ \ \b)}_{\ \ \ \c} \el
  g^{\a\b} R_{\a\b} &\rightarrow& g^{\a\b} R_{\a\b} + g^{\a\b} \nabla_\c \Sigma^{\c}_{\ \ \a\b} - g^{\a\b} \nabla_\a \Sigma^{\c}_{\ \ \c\b} + \Sigma^{\a\b}_{\ \ \ \b } \Sigma^{\c}_{\ \c\a} - \Sigma_{\a\b\c} \Sigma^{\c\a\b} - T_{\a\b\c} \Sigma^{\c\a\b}
  \el
  &=& R -  \left( \ha Q_\a + T_\a \right) ( \Sigma^{\a\b}_{\ \ \ \b } - \Sigma_{\b}^{\ \b\a} ) - T_{\a\b\c} \Sigma^{\c\a\b} + Q_{\c\a\b} \Sigma^{\c\a\b} - \hat Q^\a \Sigma^{\b}_{\ \b\a} \el
  && + \Sigma^{\a\b}_{\ \ \ \b} \Sigma^{\c}_{\ \c\a} - \Sigma_{\a\b\c} \Sigma^{\c\a\b} + \mathring\nabla_\a ( \Sigma^{\a\b}_{\ \ \ \b } - \Sigma_{\b}^{\ \b\a} ) \ .
\eea
where we have applied \re{boundary}. In the unconstrained case where torsion is present, this transformation generates a term proportional to $T^\a\pat_\a h$ in the action. As we have assumed that the torsion enters only via the connection that in turn enters only via the Riemann tensor and covariant derivatives, we have no such term in the action to begin with. Demanding that such terms are not generated requires $T_\a ( \Sigma^{\a\b}_{\ \ \ \b } - \Sigma_{\b}^{\ \b\a} ) + T_{\a\b\c} \Sigma^{\c\a\b}=0$, which according to \re{trans} is equivalent to $\Sigma_3=\Sigma_1-\Sigma_2$. Then we also have $\Sigma^{\a\b}_{\ \ \ \b } - \Sigma_{\b}^{\ \b\a}=0$, and the boundary term in \re{shiftG} vanishes. (If we had a $T^\a\pat_\a h$ term in the original action, we would fix $\Sigma_3$ by demanding that the new term cancels it.)
In the constrained case where the torsion is put to zero a priori, we instead have $\Sigma_3=0$.\footnote{In this case we are also left with a boundary term, which we discard. We used the principle of not adding or subtracting boundary terms when writing down the action. However, the field transformation is just a calculational device. We would get the same result by solving the field equations without making field redefinitions.}

Redefining the functions of $h$ in the action to absorb these changes, for the scalar field transformation $h\rightarrow\chi$ we get
\bea \label{funh}
  K &\rightarrow& \tilde K = \left(\frac{\rmd h}{\rmd\chi}\right)^2 K \el
  A_i &\rightarrow& \tilde A_i = \frac{\rmd h}{\rmd\chi} A_i \ .
\eea
For the metric transformation $g_{\a\b}\rightarrow\Omega^{-1}g_{\a\b}$ applied to the action \re{actionJg} we get
\bea \label{fung}
  F &\rightarrow& \tilde F = \Omega^{-1} F \el
  K &\rightarrow& \tilde K = \Omega^{-1} [ K - 2 ( A_1 + 4 A_2 ) \omega' - 2 ( 4 B_1 + B_2 + B_3 + 16 B_4 + 4 B_5 ) \omega'{}^2 ]
 \el
  V &\rightarrow& \tilde V = \Omega^{-2} V \el
  A_1 &\rightarrow& \tilde A_1 = \Omega^{-1} [ A_1 + ( 2 B_2 + 2 B_3 + 4 B_5 ) \omega' ] \el
  A_2 &\rightarrow& \tilde A_2 = \Omega^{-1} [ A_2 + ( 2 B_1 + 8 B_4 + B_5 ) \omega' ] \el
  B_i &\rightarrow& \tilde B_i = \Omega^{-1} B_i \ .
\eea
Finally, the connection transformation $\Gamma^{\c}_{\a\b}\rightarrow\Gamma^{\c}_{\a\b} + \Sigma^{\c}_{\ \a\b}$ gives, using the decomposition \re{transG},
\bea \label{funG}
  K \rightarrow \tilde K &=& K - 3 F ( \Sigma_1^2 + \Sigma_2^2 + \Sigma_3^2 + 4 \Sigma_1 \Sigma_2 + 2 \Sigma_2 \Sigma_3 + 4 \Sigma_3 \Sigma_1 ) \el
  &&- 2 A_1 ( 5 \Sigma_1 + 7 \Sigma_2 + 3 \Sigma_3 ) - 4 A_2 ( \Sigma_1 + 5 \Sigma_2 - 3 \Sigma_3 ) \el
 && - 4 B_1 ( 5 \Sigma_1^2 + 17 \Sigma_2^2 + 9 \Sigma_3^2 + 14 \Sigma_1 \Sigma_2 - 6 \Sigma_2 \Sigma_3 + 6 \Sigma_3 \Sigma_1 ) \el
 && - 2 B_2 ( 7 \Sigma_1^2 + 31 \Sigma_2^2 - 9 \Sigma_3^2 + 34 \Sigma_1 \Sigma_2 + 6 \Sigma_2 \Sigma_3 - 6 \Sigma_3 \Sigma_1 ) \el
&& - 2 B_3 ( 5 \Sigma_1 + 7 \Sigma_2 + 3 \Sigma_3 )^2 - 8 B_4 ( \Sigma_1 + 5 \Sigma_2 - 3 \Sigma_3 )^2 \el
  && - 4 B_5 ( 5 \Sigma_1^2 + 35 \Sigma_2^2 - 9 \Sigma_3^2+ 32 \Sigma_1 \Sigma_2 - 6 \Sigma_2 \Sigma_3 - 12 \Sigma_3 \Sigma_1 ) \el
  A_1 \rightarrow \tilde A_1 &=& A_1 + \frac{1}{2} F ( \Sigma_1 + 3 \Sigma_2 + 3 \Sigma_3 ) + 4 B_1 ( \Sigma_1 + \Sigma_2 + \Sigma_3 ) + 2 B_2 ( \Sigma_1 + 3 \Sigma_2 - \Sigma_3 ) \el
  && + 2 B_3 ( 5 \Sigma_1 + 7 \Sigma_2 + 3 \Sigma_3 ) + 2 B_5 ( \Sigma_1 + 5 \Sigma_2 - 3 \Sigma_3 ) \el
  A_2 \rightarrow \tilde A_2 &=& A_2 + \frac{1}{4} F ( \Sigma_1 - 3 \Sigma_2 - 3 \Sigma_3 ) + 4 B_1 ( \Sigma_2 - \Sigma_3 ) + 2 B_2 ( \Sigma_1 + \Sigma_2 + \Sigma_3 ) \el
  && + 4 B_4 ( \Sigma_1 + 5 \Sigma_2 - 3 \Sigma_3 ) + B_5 ( 5 \Sigma_1 + 7 \Sigma_2 + 3 \Sigma_3 ) \ .
\eea

\para{Unconstrained case.}

In the case when the torsion is left free to be determined by the field equations and we have no $T^\a \pat_\a h$ terms in the original action, we have $\Sigma_3=\Sigma_1-\Sigma_2$. Using \re{funG} and setting $\tilde A_i=0$ then fixes $\Sigma_1, \Sigma_2$ as
\bea \label{Sigmator} 
  \Sigma_1 &=& \frac{ - A_1 ( 2 B_1 + 8 B_4 + B_5 ) + 2 A_2 ( B_2 + B_3 + 2 B_5 ) }{M} \el
  \Sigma_2 &=& \frac{ A_1 ( - 8 B_1 + 8 B_2 - 16 B_4 + 16 B_5 - F ) - 4 A_2 ( 4 B_1 + 8 B_3 - 2 B_5 + F ) }{8 M} \ ,
\eea
where $M$ is defined in \re{qttor}.

We can now take $\Omega=F$ to set $\tilde F=\Mpl^2$ (and choose units such that $\Mpl=1$) and then choose $\Sigma_1, \Sigma_2$ to set $\tilde A_i=0$. According to \re{fung} and \re{Sigmator} we get
\bea \label{Sigmator2}
  \Sigma_1 &=& \{ - [ A_1 + ( 2 B_2 + 2 B_3 + 4 B_5 ) \omega' ] ( 2 B_1 + 8 B_4 + B_5 ) \el
  && + 2 [ A_2 + ( 2 B_1 + 8 B_4 + B_5 ) \omega' ] ( B_2 + B_3 + 2 B_5 ) \} M^{-1} \el
  \Sigma_2 &=&  \{ [ A_1 + ( 2 B_2 + 2 B_3 + 4 B_5 ) \omega' ] ( - 8 B_1 + 8 B_2 - 16 B_4 + 16 B_5 - F ) \el
  && - 4 [ A_2 + ( 2 B_1 + 8 B_4 + B_5 ) \omega' ] ( 4 B_1 + 8 B_3 - 2 B_5 + F ) \}  ( 8 M)^{-1} \ .
\eea

It is straightforward to check that this shift gives the same result for the connection as solving for $Q_{\a\b\c}$ directly from \re{qteq} and applying \re{L}, up to terms proportional to $F'$, which have to be accounted for by a conformal transformation of the metric. The correspondence is
\bea
  \Sigma_1 &=& \ha q_1 - q_2 - t + \frac{F'}{2 F} \el
  \Sigma_2 &=& - \ha q_1 + \ha t - \frac{F'}{4 F} \ .
\eea

The overall transformation of the kinetic term is, according to \re{fung} and \re{funG} and taking into account $\Sigma_3=\Sigma_1-\Sigma_2$,
\bea \label{Ktor}
  K &\rightarrow& F^{-1} \{ K - 2 ( A_1 + 4 A_2 ) \omega' - 2 ( 4 B_1 + B_2 + B_3 + 16 B_4 + 4 B_5 ) \omega'{}^2 \el
  && -18 F \Sigma_1^2 - 8 [ A_1 + ( 2 B_2 + 2 B_3 + 4 B_5 ) \omega' ] ( 2 \Sigma_1 + \Sigma_2 ) \el
  && + [ A_2 + ( 2 B_1 + 8 B_4 + B_5 ) \omega' ] ( 8 \Sigma_1 - 32 \Sigma_2 ) + B_1 ( - 80 \Sigma_1^2 - 128 \Sigma_2^2 + 64 \Sigma_1 \Sigma_2 ) \el
  && + B_2 ( 16 \Sigma_1^2 - 32 \Sigma_2^2 - 128 \Sigma_1 \Sigma_2 ) - B_3 ( 128 \Sigma_1^2 + 32 \Sigma_2^2 + 128 \Sigma_1 \Sigma_2 ) \el
  && + B_4 ( - 32 \Sigma_1^2 - 512 \Sigma_2^2 + 256 \Sigma_1 \Sigma_2 ) + B_5 ( 64 \Sigma_1^2 - 128 \Sigma_2^2 - 224 \Sigma_1 \Sigma_2 ) \} \ ,
\eea
with $\Sigma_1, \Sigma_2$ given by \re{Sigmator2}. 

As we have discussed, with $\tilde F'=0$ and $\tilde A_i=0$, the connection equation of motion \re{Gammaeq} gives $Q_{\a\b\c}=T_{\a\b\c}=0$ (because the $B_i$ terms break the projective invariance), so $\Gamma^{\c}_{\a\b}$ becomes the Levi--Civita connection, and the system reduces to metric gravity with a minimally coupled scalar field $h$. We can then choose $\chi$ to set $\tilde K=1$, though in some cases it may be more convenient to keep using $h$ and retain a non-canonical kinetic term, if $\frac{\rmd\chi}{\rmd h}=\pm\sqrt{K(h)}$ cannot be integrated in closed form. In terms of the action, we have
\bea
  \label{actiontrans} S &=& \int\rmd^4 x \sqrt{-g} \left[ \ha g^{\a\b} R_{\a\b}(\Gamma, \pat\Gamma) - \ha K(h) g^{\a\b} \nabla_\a h \nabla_\b h - \frac{V(h)}{F(h)^2} \right] \el
  &=&  \int\rmd^4 x \sqrt{-g} \left[ \ha g^{\a\b} R_{\a\b}(\Gamma, \pat\Gamma) - \ha g^{\a\b} \nabla_\a \chi \nabla_\b \chi - U(\chi) \right]  \ ,
\eea
where $K(h)$ is given by \re{Ktor} and $U(\chi) \equiv V[h(\chi)]/F[h(\chi)]^2$.

\para{Zero torsion.}

When torsion is put to zero a priori, we have $\Sigma_3=0$, so according to \re{funG} setting $\tilde A_i=0$ fixes $\Sigma_1, \Sigma_2$ as
\bea \label{Sigmanotor}
  \Sigma_1 &=& \frac{ - A_1 ( 16 B_1 + 8 B_2 + 80 B_4 + 28 B_5 - 3 F ) + 2 A_2 ( 8 B_1 + 12 B_2 + 28 B_3 + 20 B_5 + 3 F ) }{N} \el
  \Sigma_2 &=& \frac{ A_1 ( 8 B_2 + 16 B_4 + 20 B_5 + F ) - 2 A_2 ( 8 B_1 + 4 B_2 + 20 B_3 + 4 B_5 + F ) }{N} \ ,
\eea
with $N$ given in \re{qtnotor}. It is again straightforward to verify that this shift gives the same result for the connection as solving for $Q_{\a\b\c}$ directly from \re{qteq} and applying \re{L}, up to terms proportional to $F'$.

Again first transforming the metric to get $\tilde F=1$ and then adjusting the connection to get $\tilde A_i=0$ we have, from \re{fung} and \re{funG},
\bea \label{Sigmanotor2}
  \Sigma_1 &=& \{ - [ A_1 + ( 2 B_2 + 2 B_3 + 4 B_5 ) \omega' ] ( 16 B_1 + 8 B_2 + 80 B_5 + 28 B_5 - 3 F ) \el
  && + 2 [ A_2 + ( 2 B_1 + 8 B_4 + B_5 ) \omega' ] ( 8 B_1 + 12 B_2 + 28 B_3 + 20 B_5 + 3 F ) \} N^{-1} \el
  \Sigma_2 &=& \{ [ A_2 + ( 2 B_1 + 8 B_4 + B_5 ) \omega' ] ( 8 B_2 + 16 B_4 + 20 B_5 + F ) \el
  && - 2 [ A_2 + ( 2 B_1 + 8 B_4 + B_5 ) \omega' ] ( 8 B_1 + 4 B_2 + 20 B_3 + 4 B_5 + F ) \} N^{-1} \ ,
\eea
Again, in the chosen field coordinates the connection is Levi--Civita, and all effects of the original non-minimal coupling and non-metricity are shifted to the kinetic term and potential of the scalar field, and the action has the form \re{actiontrans}. The kinetic term reads, from \re{transg} and \re{transG},
\bea \label{Knotor}
  K &\rightarrow& F^{-1} \{ K - 2 ( A_1 + 4 A_2 ) \omega' - 2 ( 4 B_1 + B_2 + B_3 + 16 B_4 + 4 B_5 ) \omega'{}^2 \el
  && - 3 F ( \Sigma_1^2 + \Sigma_2^2 + 4 \Sigma_1 \Sigma_2 ) - 2 [ A_1 + ( 2 B_2 + 2 B_3 + 4 B_5 ) \omega' ] ( 5 \Sigma_1 + 7 \Sigma_2 ) \el
  && - 4 [ A_2 + ( 2 B_1 + 8 B_4 + B_5 ) \omega' ] ( \Sigma_1 + 5 \Sigma_2 ) \el
  && - 4 B_1 ( 5 \Sigma_1^2 + 17 \Sigma_2^2 + 14 \Sigma_1 \Sigma_2 )  - 2 B_2 ( 7 \Sigma_1^2 + 31 \Sigma_2^2 + 34 \Sigma_1 \Sigma_2 ) - 2 B_3 ( 5 \Sigma_1 + 7 \Sigma_2 )^2 \el
  && - 2 B_3 ( 5 \Sigma_1 + 7 \Sigma_2 )^2 - 8 B_4 ( \Sigma_1 + 5 \Sigma_2 )^2 - 4 B_5 ( 5 \Sigma_1^2 + 35 \Sigma_2^2 + 32 \Sigma_1 \Sigma_2 ) \} \ ,
\eea
with $\Sigma_1, \Sigma_2$ given by \re{Sigmanotor2}. 

\section{The Higgs case} \label{sec:Higgs}

\subsection{The potential and CMB observables} \label{sec:pot}

\para{Structure of the potential and the kinetic term.}

We have fully mapped the effects of the non-minimal coupling and non-metricity term to the kinetic term and the potential in the scalar sector, so their effect can be analysed with the usual inflationary vocabulary. Let us now specialise to the case of Higgs inflation and consider what kind of phenomenology the terms can lead to.

We identify $h$ with the SM background Higgs field. As the Higgs is part of a doublet, it only appears in even powers in the action. Restricting to terms of up to dimension 4, we have
\bea \label{Higgsfun}
  K = k \ , \qquad F = f + \xi h^2 \ , \qquad A_i = a_i h \ , \qquad B_i = b_{i0} + b_{i1} h^2 \ ,
\eea
where $k, f, \xi, a_i,  b_{i0}, b_{i1}$ are constants. Note that $f$ does not have to be close to $\Mpl^2=1$, since the Planck scale is defined in terms of gravity analysed in the Einstein frame. The SM Higgs potential is
\bea \label{V}
  V(h) = \frac{1}{4} \lambda (h^2-v^2)^2 \ .
\eea

The only thing that changes the potential as a function of $h$ is the non-minimal coupling $F$, which (for $\xi\neq0$) flattens the transformed potential for large values of $h$ as usual in Higgs inflation,
\bea
  U(\chi) \equiv \frac{V[h(\chi)]}{\Omega[h(\chi)]^2} = \frac{V[h(\chi)]}{F[h(\chi)]^2} = \frac{\lambda}{4} \frac{(h^2-v^2)^2}{(f + \xi h^2)^2} \simeq \frac{\lambda}{4 \xi^2} \left( 1 - \frac{2 f}{\xi h^2} \right) \ .
\eea
However, the potential as a function of $\chi$ is affected also by the redefinition of the kinetic term. The kinetic function has the same qualitative structure in the unconstrained case \re{Ktor} and in the zero torsion case \re{Knotor}. With the functions \re{Higgsfun}, the kinetic term in both cases reads
\bea \label{Kh}
  K(h) &=& \frac{k}{f + \xi h^2} + \frac{ h^2 }{ ( f + \xi h^2 )^3 ( O_0 + O_1 h^2 + O_2 h^4 )^2 } \sum_{n=0}^{5} K_n h^{2n} \ ,
\eea
where $K_n$ and the coefficients of $O(h)\equiv O_0+O_1 h^2+O_2 h^4$ are complicated polynomials of the constants defined in \re{Higgsfun}. The function $K(h)$ is determined by and \re{Sigmator2} and \re{Ktor} (with torsion) or \re{Sigmanotor2} and \re{Knotor} (no torsion). In the case with torsion, $O(h)=M(h)$ given in \re{qttor}, and in the case with no torsion, $O(h)=N(h)$ given in \re{qtnotor}.

The field transformation to the canonically normalised field $\chi$ is given by
\bea \label{chih}
  \frac{\rmd\chi}{\rmd h} = \pm \sqrt{K(h)} \ .
\eea

In the small field limit $h\ll1$, the second term in \re{Kh} does not contribute, so \re{chih} gives $\chi=\sqrt{k/f} h$, and the potential reduces to $U(\chi)=V(\sqrt{f/k}\chi)/f^2$. The parameters of the Higgs potential \re{V} written in terms of the canonical field become $\lambda\rightarrow\lambda/k^2, v\rightarrow\sqrt{k/f} v$. The Higgs mass correspondingly changes as $m_h=\sqrt{2\lambda}v\rightarrow m_h/\sqrt{k f}$. Starting from the original action, we can therefore scale the quartic coupling and the Higgs mass at will. In particular, a large Higgs mass can be brought down to the EW scale by choosing a large $f$. Restoring dimensions, we have $F=M^2+\xi h^2$ and $\Omega=\frac{\Mpl^2}{M^2+\xi h^2}$. In the limit of small $h$, the transformation to the minimally coupled canonical field reduces to the field-independent redefinition $m_h\rightarrow\frac{1}{\sqrt{k}}\frac{M}{\Mpl}m_h$, so the hierachy between $m_h$ and $\Mpl$ can be transformed into the question why $M$ is much smaller than $\Mpl$.

\para{CMB observables.}

The observational constraints from cosmic microwave background (CMB) data on the scalar perturbation amplitude $A_s$, scalar spectral index $n_s$ and tensor-to-scalar ratio $r$ are (assuming no running of the spectral index) \cite{Planck2018}
\bea \label{CMB}
  24 \pi^2 A_s &=& \frac{U}{\epsilon} = ( 4.97 \pm 0.07 ) \times 10^{-7} \el
   n_s &=& 1 + 6 \epsilon - 2 \eta = 0.9653 \pm 0.0041 \el
   r &=& 16 \epsilon < 0.07 \ ,
\eea
where the slow-roll parameters are $\epsilon\equiv\ha (U'/U)^2, \eta\equiv U''/U$.

The inflationary behaviour is determined by the shape of $U(h)$ and $K(h)$. Let us consider some possibilities and compare to the observed values \re{CMB}.

\subsection{Types of inflationary potentials}

\para{Large-field case with $\xi\neq0$.}

If the non-minimal coupling is non-zero, the potential $U(h)$ is asymptotically flat in the limit of large $h$, and the behaviour of $K(h)$ in this limit determines how rapidly the plateau is approached.\footnote{How large $h$ has to be depends on the value of $\xi$ and the coefficients appearing in $K(h)$. As there are many terms, the numerical factors multiplying $h$ can be large, so large field does not necessarily mean $h\gg1$.} The function $K(h)$ is the ratio of two polynomials, the numerator being of order 12 and the denominator of order 14. Thus, in the generic case when neither of the leading order coefficients is tiny, $K(h)$ is proportional to $h^{-2}$ in the large field limit. This leads to an exponential potential for $\chi$ as in usual Higgs inflation. In the case $A_i=B_i=0$, we have $K=k/(\xi h^2)$, and for $k=1$ we get $h=\frac{1}{2\sqrt{\xi}}e^{\sqrt{\xi}\chi}$ and $U=\frac{\lambda}{4\xi^2}(1-2e^{-2\xi\chi})$ \cite{Bauer:2008}. In the general case, the coefficient of the leading contribution is a complicated polynomial of the constants defined in \re{Higgsfun}. In the limit $h\gg1,\xi\gg1$ the kinetic function reduces to $K\simeq[ k + \frac{ ( a_1 + 4 a_2 )^2 }{8 b_{11} + 2 b_{21} + 2 b_{31} + 32 b_{41} + 8 b_{51}} ]/(\xi h^2)$ in the unconstrained case and to $K\simeq( k - 4 a_1 - 16 a_2 - 32 b_{11} - 8 b_{21} - 8 b_{31} - 128 b_{41} - 32 b_{51} )/(\xi h^2)$ in the zero torsion case. In the zero torsion case this corresponds to the replacement $\xi\rightarrow\xi/(k - 4 a_1 - 16 a_2 - 32 b_{11} - 8 b_{21} - 8 b_{31} - 128 b_{41} - 32 b_{51})$ in the exponential potential, and correspondingly in the unconstrained case. In the tree-level Palatini case with the exponential potential the normalisation of the perturbations gives $\lambda/\xi=10^{-10}$, requiring $\xi=10^9$ for $\lambda=0.1$. (When loop corrections change the potential, $\xi$ can be orders of magnitude smaller \cite{Rasanen:2017, Enckell:2018a, Rasanen:2018a}.) If the coefficients dividing $\xi$ are of order unity, $\xi$ is changed by at most around two orders of magnitude, unless they conspire so that the sum is much smaller than the individual terms. Absent such tuning, when the leading terms of the polynomials in the numerator and denominator of $K(h)$ dominate, the situation is not much changed from the case with $A_i=B_i=0$.

Different behaviour can be obtained by choosing the coefficients so that the leading term of the numerator or the denominator vanishes. For example, with the choice $k=1, f=1, \xi=2\times10^5$ and $a_i=60, b_{i0}=b_{i1}=-2.3747\times10^4$ (with torsion) or $a_i=14.5063, b_{i0}=b_{i1}=-1.39$ (no torsion), the $h^{12}$ term in the numerator in suppressed relative to the $h^{10}$ term, so for large $h$ we have $K(h)\simeq \kappa_1 h^{-4}$ with $\kappa_1=10^{-3}$. The relation \re{chih} then gives $\chi=\sqrt{\kappa_1} h^{-1}$, and the potential becomes
\bea \label{hilltop}
  U(\chi) \simeq \frac{\lambda}{4\xi^2} \left( 1 - \frac{2}{\xi\kappa_1} \chi^2 \right) \ ,
\eea
with $\chi\ll1$. This is a small-field potential of the hilltop type. We have chosen the parameters (taking again $\lambda=0.1$) to reproduce the amplitude $A_s$ and the spectral index $n_s$ in agreement with the observations \re{CMB}, taking the CMB pivot scale to correspond to $\chi=0.06$. We did not check the constraint of obtaining the right number of e-folds. For this example, the tensor-to-scalar ratio is $r=1\times10^{-5}$. The value of $\xi$ is four orders of magnitude smaller than in the case with only the non-minimal coupling $F$, because now $U(\chi)$ is less flat, so the slow-roll parameter $\epsilon$ is larger; recall that $24\pi^2 A_s=U/\epsilon=16 U/r$. Hilltop-type Higgs inflation has previously been considered with loop corrections \cite{Fumagalli:2016lls, Rasanen:2017, Enckell:2018a}, here we get similar behaviour at the classical level. As the polynomials in $K(h)$ are of high order and the coefficients are large, canceling the leading terms requires a lot of tuning. This is also true for an observationally viable hilltop generated by quantum corrections.

Alternatively, we can choose the coefficients \re{Higgsfun} so that $O_2=0$ and the two leading terms of the denominator in \re{Kh} vanish.\footnote{It is not possible to make just the leading term to vanish identically, although we could try to tune $O_1$ and $O_2$ so that the leading term is strongly subdominant to the next-to-leading term. This would lead to a potential of the form $U(\chi)\propto1-\kappa_2\chi^{-2}$, which gives a viable inflationary model.}
In this case $K(h)\simeq\kappa_3 h^2$ for large $h$, so \re{chih} gives $\chi=\frac{\sqrt{\kappa_3}}{3} h^3$, and the potential becomes
\bea
  U(\chi) \simeq \frac{\lambda}{4\xi^2} \left( 1 - \frac{2 f}{\xi \kappa_3} \chi^{-2/3} \right) \ .
\eea
In the case with torsion, this is not possible, as all of the leading terms of the numerator are proportional to $M^2$, so they will also be suppressed. In the case without torsion, this case can be realised for example by choosing $\xi=11$, $a_1=315, a_2=289, b_{i0}=1, b_{i1}=1.3955767262643881$; then $\chi=8$ at the pivot scale gives $A_s$ and $n_s$ in agreement with the central values \re{CMB}, and $r=0.04$. Again, we did not check the number of e-folds. For these coefficients, the $1/h^2$ term and other higher order terms can contribute to $K(h)$, and might have to be suppressed by further tuning of coefficients.

\para{Large-field case with $\xi=0$.}

Let us now consider the case without non-minimal coupling.\footnote{In general, $\xi$ is generated by quantum corrections \cite{Callan:1970ze}, and it will run with scale, and thus cannot be zero on all scales. However, it can be negligibly small in the inflationary region.} Now $U$ is not asymptotically flat. We have $K_4=K_5=0$, so the numerator and denominator of $K(h)$ are both polynomials of order 8, with complicated coefficients. If coefficients are not tuned, in the large field limit the field transformation \re{chih} thus simply rescales the field by a constant, so we have $U\propto\chi^4$. The rescaling could be used to make the value of the effective quartic coupling compatible with the constraint from the perturbation amplitude, but $n_s$ would anyway be below and $r$ above the observational bound \re{CMB} \cite{Planck2018}.

If we choose the leading term of the numerator to vanish, we have $K(h)=\kappa_4 h^{-2}$ for large $h$. This gives $h\propto e^{\chi/\sqrt{\kappa_4}}$, so $U\propto e^{4\chi/\sqrt{\kappa_4}}$. Such a potential gives $r$ in excess of the observational upper bound.

If we instead choose the two leading terms of the denominator to vanish, we have $K(h)\propto h^4$ for large $h$, leading to $\chi\propto h^3$ and $U\propto \chi^{4/3}$, which gives $r\gtrsim0.1$, taking into account the constraint \re{CMB} on $n_s$, and is thus excluded. Suppressing only the first leading term would give $K(h)\simeq h^2$, $\chi\propto h^2$ and $U\propto\chi^2$, which would also give to a too large $r$.

\para{$\a$-attractor.}

We can have inflation at intermediate values of $h$ if the potential is very flat there. There are two possibilities. We can have an $\a$-attractor, where the denominator of $K(h)$ is zero \cite{Ferrara:2013, Kallosh:2013, Galante:2014},\footnote{We can also have an $\a$-attractor at large field values. Usual Higgs inflation in the large $h$ limit can be understood as an $\a$-attractor model \cite{Galante:2014, Rubio:2018}.} or we can have an inflection point (or a near-inflection point) where $K(h)$ remains regular. In the context of Higgs inflation an inflection point is usually also an extremum (or nearly so), and it is then called a critical point \cite{Allison:2013uaa, Bezrukov:2014bra, Hamada:2014iga, Bezrukov:2014ipa, Rubio:2015zia, Fumagalli:2016lls, Enckell:2016xse, Bezrukov:2017dyv, Rasanen:2017, Ezquiaga:2017fvi, Rasanen:2018a}.

Let us first consider the $\a$-attractor case. The kinetic function $K(h)$ has a zero in the denominator at $h_0$ if the functions $F, B_i$ are chosen so that $O(h_0)=0$. This produces an inflationary plateau if the numerator is positive at $h_0$. In the case of Higgs inflation \re{Higgsfun}, the parameters $f,\xi,b_{i0}, b_{i1}$ can be chosen to make a pole whose order varies from $(h-h_0)^{-1}$ to $(h-h_0)^{-4}$. In order to have a viable model, we must also demand that $K(h)>0$ for $0\leq h<h_0$.
These conditions are easy to satisfy. For example, in the case with torsion the choice of parameters
$k=1, f=1, \xi=0, a_i=a, b_{10}=1, b_{11}=3\times10^4, b_{j0}=-1, b_{j1}=0$ (with $j=2,3,4,5$) with a non-zero constant $a$, gives a first order pole, $K(h)\simeq\kappa_5 (h-h_0)^{-1}$ at $h_0=0.01$ with $\kappa_5=1\times10^{-6} a^2$. (The positivity of $K(h)$ constrains the possible values of $a$.) In the case with no torsion we can choose $k=1, f=1, \xi=0, a_i=a, b_{10}=b_{50}=1, b_{j0}=-1, b_{11}=1, b_{51}=5000, b_{j1}=0$ (with $j=2,3,4$) to get a second order pole at $h_0=0.01$ with $\kappa_5=2\times10^{-8} a^2$. In the limit of large number of e-folds $N$, the second order pole gives \cite{Galante:2014}
\bea
  n_s = 1 - \frac{2}{N} \ , \quad r = \frac{8\kappa_5}{N^2} \ .
\eea
The spectral index is the same as in the usual Higgs inflation case (and thus agrees with observations for $N=50$, the usual number of e-folds for the case without the extra terms \cite{Rasanen:2017, Figueroa:2009, Figueroa:2015, Repond:2016} -- though see \cite{Ema:2016, DeCross:2016}), whereas the tensor-to-scalar ratio can be adjusted at will. The normalisation of the perturbations \re{CMB} implies $\lambda=10^{-7} r h_0^{-4}$, which for the above examples gives $r=0.08\lambda$. However, the parameters could be adjusted to change $h_0$ to make $r$ smaller or larger.

\para{Inflection point.}

We can also choose parameters to produce an inflection point, where $\rmd^2 U/\rmd\chi^2=0$ and $K(h)$ remains regular and positive. The important difference to an $\a$-attractor is that the field can roll through an inflection point. In the context of Higgs inflation, inflection points from quantum corrections have been considered to get large values of $r$ \cite{Allison:2013uaa, Bezrukov:2014bra, Hamada:2014iga, Bezrukov:2014ipa, Rubio:2015zia, Fumagalli:2016lls, Enckell:2016xse, Bezrukov:2017dyv, Rasanen:2017} or an enhancement in the spectrum to produce primordial black holes as dark matter candidates \cite{Ezquiaga:2017fvi, Rasanen:2018a}.

It is easy to tune the parameters to make a regular inflection point. For example, in the case with torsion, $k=1, f=1, \xi=0, a_1=1, a_2=-100, b_{20}=-1, b_{j0}=1, b_{i1}=-1$, (with $j=1,3,4,5$) gives an inflection point at $h=0.77$. In the case without torsion, $k=1, f=1, \xi=0, a_1=1, a_2=100, b_{20}=-1, b_{j0}=1, b_{i1}=1$ (with $j=1,3,4,5$) gives an inflection point at $h=0.04$.

We cannot get $\rmd U/\rmd\chi=0$ exactly (unless $\xi<0$), because $U$ is monotonic as a function of $h$, and while field redefinitions can change the curvature of the potential, they do not affect the existence or locations of its extrema. Therefore we cannot obtain hillclimbing inflation \cite{Jinno:2017a, Jinno:2017b}, nor have an exact hilltop \cite{Fumagalli:2016lls, Rasanen:2017, Enckell:2018a} or an exact critical point. However, it may be possible to suppress $\rmd U/\rmd\chi$ by tuning the coefficients so as to get a near-critical point or a potential that looks similar to the vicinity of a hilltop. We leave investigation of such possibilities and detailed comparison to observations, including checking the right number of e-folds, for future work.

\section{Discussion} \label{sec:disc}

\para{Fermions and gauge bosons.}

Let us now look at fermions and gauge bosons. Considering only terms with up to a total of two derivatives, the only new fermionic contributions are
\bea
  S_{\mathrm{f}} &=& \int\rmd^4 x \sqrt{-g} \left[ D_1(h) \bar\psi\gamma^\a Q_\a \psi + D_2(h) \bar\psi\gamma^\a \hat Q_\a \psi \right] \ .
\eea
With the solution \re{Qsol} for $Q_{\a\b\c}$, we will end up with a term of the form $g(h)\bar\psi\gamma^\a \pat_\a h \psi$. In the SM, where we have a $U(1)$ gauge field, such a term corresponds to a gauge transformation, and thus does not lead to any physical effect such as generation of gravitational waves, unlike the corresponding pseudoscalar term.\footnote{Thanks to Lorenzo Sorbo for pointing this out.}
Fermions also contribute to the equation of motion of $\Gamma^{\c}_{\a\b}$ by generating a spin current via their coupling to the connection \cite{Hehl:1978, Hehl:1981, Sotiriou:2006}. This is not important during inflation when the scalar field dominates, and while fermions become significant during reheating, there is no overall spin order, so the spin current is expected to be highly suppressed.

As for gauge fields, if we restrict to a total of two derivatives, there are no new terms involving $Q_{\a\b\c}$. If we allow terms with more than two derivatives in total (as long as they are not acting on the same field), at leading order in the derivatives we have
\bea
  S_{\mathrm{g}} &=& \int\rmd^4 x \sqrt{-g} \left[ E_1(h) F_{\a\b} Q^{\a\b\c} Q_\c + E_2(h) F_{\a\b} Q^{\a\b\c} \hat Q_\c + E_3(h) F_{\a\b} Q^\a \hat Q^\b \right. \el
  && \left. + E_4(h) \epsilon^{\a\b\c\d} F_{\a\b} Q_{\c\d}^{\ \ \ \epsilon} Q_\epsilon + E_5(h) \epsilon^{\a\b\c\d} F_{\a\b} Q_{\c\d}^{\ \ \ \epsilon} \hat Q_\epsilon + E_6(h) \epsilon^{\a\b\c\d} F_{\a\b} Q_\c \hat Q_\d \right] \ ,
\eea
where $F_{\a\b}$ is an Abelian field strength. In the case of the SM, $F_{\a\b}$ is the $U(1)_Y$ gauge field strength, and these terms terms break the conformal symmetry of the gauge field, so they might at first sight seem like candidates for inflationary magnetogenesis \cite{Turner:1987bw, Durrer:2013pga}. However, when $Q_{\a\b\c}$ is sourced only by a single scalar field, all of these contributions vanish in the equations of motion due to the symmetry of the $Q_{\a\b\c}$ contributions paired with the antisymmetry of $F_{\a\b}$ or $\epsilon_{\a\b\c\d}$. When fermions and gauge bosons contribute significantly during preheating, this could change. The situation would also be different in models where more than one scalar field contributes, as then $Q^\a$ and $\hat Q^\a$ are in general not aligned.

\para{Higher order $Q_{\a\b\c}$ terms.}

We have allowed at most two derivatives in the action, but as the partial derivatives of the metric vanish (by pairing with the Levi--Civita connection in the equations of motion), higher powers of $Q_{\a\b\c}$ do not lead to higher order of the equations of motion. The restriction therefore seems arbitrary. Keeping to terms of up to dimension 4, there are no new terms mixing the derivatives the metric with the scalar field, fermions or gauge bosons, nor are there any new terms of order $(\nabla g)^3$. The only new terms are of order $(\nabla g)^4$, which do not couple directly to the scalar field. There are dozens of different contractions, but they all contribute in essentially the same way to the connection equation of motion \re{Gammaeq}, changing the equation from first to third order in $Q_{\a\b\c}$. The tensor structure of the solution \re{Qsol} will remain the same, but the functions $q_i$ and $t$ will depend on $g^{\a\b}\pat_\a h \pat_\b h$ in addition to $h$ via equations of the form (denoting $t\equiv q_3$) $q_k [ P_k^{(1)}(h) + \sum_{i,j=1}^{3} P^{(2)}_{kij}(h) q_i q_j g^{\a\b}\pat_\a h \pat_\b h ] = P^{(3)}_{k}(h)$, where $P_k^{(1)}, P^{(2)}_{kij}, P_k^{(3)}$ are combinations of the functions of $h$ appearing in the action. The functions $\Sigma_i$ will then depend on $g^{\a\b} \pat_\a h \pat_\b h$, and so will the kinetic function $K$. Therefore, the equation that relates $h$ to the minimally coupled scalar field $\chi$ with a canonical kinetic term will also depend on $g^{\a\b}\pat_\a h \pat_\b h$. During slow-roll inflation and in the long-wavelength limit, the derivative contribution is suppressed, so this is expected to have little effect on potential-driven slow-roll inflation. However, such terms could enable inflation driven by kinetic terms \cite{ArmendarizPicon:1999}. One issue is that as the equations are third order, they could have homogeneous solutions, meaning that the equations of motion do not completely determine the connection. Avoiding such ambiguity would translate into a constraint on the terms allowed in the action.

\section{Conclusions} \label{sec:conc}

\para{The gravity track of Higgs inflation.}

We have considered scalar field inflation in the Palatini formulation of general relativity, where the connection is an independent variable. The covariant derivative of the metric (\ie the non-metricity) is then non-zero, and it can couple to the scalar field. Assuming that the connection enters only via the Riemann tensor and the covariant derivative, we have written down all coupling terms with at most two derivatives such that the equations of motion follow from the bulk action without needing to add boundary terms. We have considered both the unconstrained case when the antisymmetric part of the connection (the torsion) is determined by the field equations and the case when it set to zero a priori.

Terms that are quadratic in the derivative of the metric are motivated by the fact that they break the projective invariance of the action and thus allow the equations of motion to fully determine the degrees of freedom in the case when the torsion is not taken to be zero. If the non-minimal coupling of the scalar field and the mixing between the kinetic terms of the metric and the scalar field are zero, the quadratic terms simply set the non-metricity and torsion to zero and make the connection Levi--Civita.
When non-minimal coupling and/or the kinetic mixing between the metric and the scalar field are non-zero, they source non-metricity and torsion. We transform to the Einstein frame, where the connection becomes Levi--Civita and the scalar field is minimally coupled to gravity, and the effects of non-metricity and torsion are mapped to the kinetic term and the potential of the scalar field.

Specialising to the Standard Model Higgs field $h$ as the inflaton (meaning that for large field values the potential is quartic and the field appears only in even powers), we study the range of inflationary models made possible by the new terms. Including only terms of up to dimension 4, the kinetic term is the ratio of two complicated polynomials of $h$, the numerator being of order 10 and the denominator of order 14.  In general, if the leading terms of the polynomials dominate at large field values, then the kinetic function is $\propto h^{-2}$ and the qualitative behaviour is the same as in usual Higgs inflation, the new terms just effectively change the value of the non-minimal coupling $\xi$. If the leading terms of either the numerator or the denominator are zero, we can get a variety of effective inflationary potentials for the canonically normalised scalar field $\chi$, such as $U\propto 1-\a \chi^2$, $U\propto 1-\a \chi^{-2/3}$, $U\propto\chi^2$, $U\propto\chi^{4/3}$, an $\a$-attractor and a potential with an inflection point. Some of these can give inflation in agreement with the observations even if the non-minimal coupling is zero. They can produce a range of values for the tensor-to-scalar ratio $r$, from too small to be detected in near-future experiments to values in excess of the current observational limits. This shows that Higgs inflation in the Palatini formulation can produce a large $r$, which has not been obtained earlier with quantum corrections nor with the inclusion of a $R^2$ term in the action \cite{Bauer:2008, Rasanen:2017, Enckell:2018a, Markkanen:2017, Rasanen:2018a}. However, we did not check the constraint on the total number of e-folds nor investigate the full parameter space, which would require a numerical study.

Getting leading terms of the polynomials to vanish requires fine-tuning (absent a symmetry that would relate the coefficients in the action to each other), just like obtaining a critical point \cite{Allison:2013uaa, Bezrukov:2014bra, Hamada:2014iga, Bezrukov:2014ipa, Rubio:2015zia, Fumagalli:2016lls, Enckell:2016xse, Bezrukov:2017dyv, Rasanen:2017, Masina:2018ejw, Salvio:2017oyf, Ezquiaga:2017fvi, Rasanen:2018a}, hilltop \cite{Fumagalli:2016lls, Rasanen:2017, Enckell:2018a} or a degenerate vacuum \cite{Jinno:2017a, Jinno:2017b} using quantum corrections. Tuning of the classical terms is also presumably unstable to quantum corrections.
The new terms could contribute to addressing the unitarity issue of Higgs inflation \cite{Barbon:2009ya, Burgess:2009ea, Hertzberg:2010dc, Bauer:2010, Bezrukov:2010jz, Bezrukov:2011sz, Calmet:2013hia, Weenink:2010rr, Prokopec:2012ug, Xianyu:2013, Prokopec:2014iya, Ren:2014, Escriva:2016cwl, Gorbunov:2018llf}, given that the non-minimal coupling can be smaller than in the usual case or even zero. They could also change renormalisation group running and affect reheating.

Extension of the original tree-level Higgs inflation proposal has been widely discussed from the point of view of quantum corrections \cite{Espinosa:2007qp, Barvinsky:2008ia, Burgess:2009ea, Popa:2010xc, DeSimone:2008ei, Bezrukov:2008ej, Bezrukov:2009db, Barvinsky:2009ii, Bezrukov:2010jz, Bezrukov:2012sa, Allison:2013uaa, Salvio:2013rja, Shaposhnikov:2013ira, George:2013iia, Calmet:2013hia, Bezrukov:2014bra, Bezrukov:2014ipa, Rubio:2015zia, George:2015nza, Saltas:2015vsc, Fumagalli:2016lls, Moss:2014, Bezrukov:2017dyv, Enckell:2016xse, Hamada:2014iga, Rasanen:2017, Enckell:2018a, Masina:2018, Markkanen:2017, Masina:2018ejw, Salvio:2017oyf, Ezquiaga:2017fvi, Rasanen:2018a, Jinno:2017b, Barbon:2009ya, Hertzberg:2010dc, Bezrukov:2011sz, Weenink:2010rr, Prokopec:2012ug, Xianyu:2013, Prokopec:2014iya, Ren:2014, Escriva:2016cwl, Bauer:2010, Calmet:2012eq, Steinwachs:2013tr, Kamenshchik:2014waa, Burns:2016ric, Hamada:2016, Karamitsos:2017elm, Karamitsos:2018lur, Postma:2014vaa, Herrero-Valea:2016jzz, Pandey:2016jmv, Pandey:2016unk}, including the generation of higher order terms in the Riemann tensor \cite{Barbon:2015, Salvio:2015kka, Salvio:2017oyf, Kaneda:2015jma, Calmet:2016fsr, Wang:2017fuy, Ema:2017rqn, Pi:2017gih, He:2018gyf, Gorbunov:2018llf, Ghilencea:2018rqg, Wang:2018, Enckell:2018b, Antoniadis:2018ywb, Gundhi:2018wyz, Antoniadis:2018yfq}.

The inclusion of kinetic terms for the metric extends the study on the parallel track of the choice of gravitational degrees of freedom. Earlier work has looked at the differences between the metric and the Palatini formulation for the same action \cite{Bauer:2008, Bauer:2010, Rasanen:2017, Enckell:2018a, Markkanen:2017, Rasanen:2018a}, whereas we have analysed the effect of terms that arise in the Palatini case and have no equivalent in the metric case. It would be interesting to extend the frame-covariant formalism developed to deal with conformal transformations of the metric \cite{Kozak:2018} and partially translations of the connection \cite{Jarv:2016sow, Karam:2017, Karamitsos:2017elm, Karamitsos:2018lur, Jarv:2014hma, Jarv:2015kga, Kuusk:2015dda} to fully cover the transformations in the space of the scalar field, the metric and the connection, and make more transparent the frame-dependent role of the non-metricity and the connection. This connects to formulating the quantum theory in terms of frame-invariant variables to settle the issue of possible non-commutation of quantisation and frame transformations \cite{Weenink:2010rr, Calmet:2012eq, Steinwachs:2013tr, Prokopec:2014iya, Kamenshchik:2014waa, Burns:2016ric, Fumagalli:2016lls, Hamada:2016, Karamitsos:2017elm, Karamitsos:2018lur, Bezrukov:2008ej, Bezrukov:2009db, Bezrukov:2010jz, Allison:2013uaa, George:2013iia, Postma:2014vaa, Prokopec:2012ug, Herrero-Valea:2016jzz, Pandey:2016jmv, Pandey:2016unk}.

From the effective theory point of view, if gravity is described by the Palatini formulation, the terms we have considered are required for not just for Higgs inflation, but for any scalar field model of inflation, and (like $R^2$ terms \cite{Enckell:2018b, Antoniadis:2018ywb, Antoniadis:2018yfq}) can significantly modify inflationary predictions.

\section*{Acknowledgements}

\noindent I thank Daniel Figueroa and Lorenzo Sorbo for helpful correspondence, Lumi-Pyry Wahlman for useful discussions and the Institute for Theoretical Physics at KU Leuven for hospitality. I thank Claire Rigouzzo and Sebastian Zell for pointing out errors in a previous version of the manuscript.

\bibliographystyle{JHEP}
\bibliography{kin}


\end{document}